**Multiple-excitation study of the double-resonance Raman bands in rhombohedral graphite**

Sergio L. L. M. Ramos[1,†], Marcos A. Pimenta[1] and Ana Champi[3,*]

[1] *Departamento de Física, Universidade Federal de Minas Gerais (UFMG), Belo Horizonte, Minas Gerais 30123-970, Brazil*

[2] *Centro de Ciências Naturais e Humanas, Universidade Federal do ABC (UFABC), Santo André, São Paulo 09210-170, Brazil*

[†] present address: *School of Chemistry, University of Manchester, Manchester M13 9PL, United Kingdom*

*Corresponding author. Tel: +55 11 973551852. E-mail: ana.champi@ufabc.edu.br

**Abstract**

The double-resonance (DR) Raman process is a signature of all $sp^2$ carbon material and provide fundamental information of the electronic structure and phonon dispersion in graphene, carbon nanotubes and different graphite-type materials. We have performed in this work the study of different DR Raman bands of rhombohedral graphite using five different excitation laser energies and obtained the dispersion of the different DR features by changing the laser energy. Results are compared with those of Bernal graphite and shows that rhombohedral graphite exhibit a richer DR Raman spectrum. For example, the 2D band of rhombohedral graphite is broader and composed by several maxima that exhibit different dispersive behavior. The occurrence of more DR conditions in rhombohedral graphite is ascribed to the fact that the volume of its Brillouin zone (BZ) is twice the volume of the Bernal BZ, allowing thus more channels for the resonance condition. The spectra of the intervalley TO-LA band of rhombohedral graphite, around 2450 cm$^{-1}$, is also broader and richer in features compared to that of Bernal graphite. Results and analysis of the spectral region 1700-1850 cm$^{-1}$, where different intravalley processes involving acoustic and optical phonons occurs, are also presented.

## 1. Introduction

In nature, a single chemical element can yield many different forms of crystalline structures, a phenomenon that is commonly referred to as allotropy. In the case of the element carbon, many allotropic forms are known to exist: diamond, graphite, fullerene, nanotubes, just to name a few; and searching for new carbon structures still remains an active field of scientific research [1–3]. The plurality of $sp^2$ carbon structures partially stems from the two-dimensional character of the atomic array that the carbon atom sets up, producing the atomically thick sheet known as graphene [4], and the numbers of ways that graphene can be stacked or rolled-up into different structures. Graphite, in its turn, represents a structure where many sheets of graphene are stacked upon each other, yielding a layered bulk material. Yet, even in this case, the way the graphene sheets are stacked is pertinent and depending on the stacking order different allotropic forms of graphite arise. The stacking configuration referred to as Bernal stacking yields graphite in its most stable and abundant form. Here, adjacent layers of stacked graphene are slightly displaced relative to each other repeating the same displacement pattern every other layer and giving rise to a stacking sequence which may be represented as ABABAB, being referred to as Bernal stacking (2H) [5]. The unit cell of 2H graphite has four inequivalent carbon atoms ($Z = 4$). Different from this, graphite can also appear with a stacking sequence of the type ABCABC in which two types of displacement patterns occur and intercalate, thereby ordering the periodicity of stacking to take place every three layers. This allotrope is designated rhombohedral graphite (3R). The primitive rhombohedral unit cell of 3R graphite has only two inequivalent carbon atoms ($Z = 2$) [5,6].

Rhombohedral graphite is reported to exist in nature in up to 30% at most, when compared to Bernal graphite in a sample of natural graphite [5,7]. Due to its metastable nature, until now it has not been possible to synthesize films or rhombohedral multi-layers of micrometric dimensions. Consequently, only few works report experimental studies in bulk or multilayers specimens. Nevertheless, rhombohedral graphite, for some time now, has sparked scientists' interests mainly for its semiconducting properties [8] and for the possible existence of superconductivity, expected to occur at reasonable temperatures as suggested by theoretical predictions [9]; experimental works further associate a possible superconducting behavior at high temperatures with interfaces of the Bernal and rhombohedral phases [10]. Bouhafs *et al*. [11]

recently attempted to synthesize rhombohedral graphite with the chemical vapor deposition method. They succeeded in producing specimens covering areas of ~50 μm$^2$ with thicknesses up to 9 monolayers; yet the crystals were characterized to composed of a mixture of alternating rhombohedral and Bernal-stacked graphitic domains. Currently most experimental works existing on the rhombohedral graphitic structure are usually reported for samples bearing thicknesses less than 6 layers [12–14]. In recent years, nevertheless, experimental and/or computational investigations focusing on thicker rhombohedral graphitic samples, *i.e.* ~ 20 layers, have begun to emerge [15–17].

Raman spectroscopy has proven to be a very useful tool to study graphite-based materials and provides different valuable information. [4,18]. For example, the presence of disorder and defects manifests itself by the appearance of disorder-induced Raman D band [19,20], and this band can also be used to distinguish between zigzag and armchair edges [21]. The stacking order can also be distinguished from the shape of the 2D band [18,22]. In the case of a graphene single layer, only one zone-center mode is allowed in the Raman spectra and gives rise to the G band, around 1580 cm$^{-1}$. For Bernal graphite, the interlayer shear mode is also Raman active and appears around 40 cm$^{-1}$ [23] while, in the case of rhombohedral graphite, the interlayer shear mode is not Raman active. The second-order spectrum of graphene systems is extremely rich due to the occurrence of the double- or triple-resonance Raman processes. These processes involve phonons within the interior of the Brillouin zone (BZ) that connect points in the two electronic valleys around the K and K' points. The intravalley process (K-K) occurs between states in the same valley and the intervalley process involves states in the K and K' valleys. The double-resonance (DR) processes also involve phonons with opposite momenta from different branches, such as the iTO, LO, LA and iTA. The intervalley DR process involving two phonons of the iTO branch gives rise to the 2D band, while the D band involves one phonon of the iTO branch and a defect, needed momentum conservation. The other features related to overtones and combination of different phonon branches are normally weak and receive different names in the literature. The DR intervalley band involving phonons of the TO and LA branches, called the G* band in Ref. [4], appears around 2450 cm$^{-1}$. The intravalley band involving phonons of the TO and ZO branches, called the M band in the literature [4], appears around 1750 cm$^{-1}$. In special, the DR process gives rise to features whose frequencies depend on excitation laser energy, since they come from phonons of dispersive branches within the interior of the BZ. Therefore, a multiple-excitation Raman investigation is needed to provide information about the electronic structure and the dispersion of phonon branches [22].

Henni *et al.* [15] concentrated characterizing the Raman features for multilayers of rhombohedral graphene/graphite by means of magneto-Raman scattering experiments. They identified a new scattering feature lying within the intermediate-frequency region, namely in between the G and 2D bands, and argued it to be an intrinsic electronic Raman scattering effect which could fingerprint the ABC stacking order. Subsequently, Torche *et al.* [16], by means of first principles calculations, considered the character of the 2D Raman band in the presence of long-range ABC stacking. Their results suggest that the different stacking structures, ABA and ABC, are characterized by different electronic band dispersions along the stacking direction $z$ and, therefore, the resonant phonon momenta that contributes to the 2D band Raman cross section is broader for the ABC stacking case. Yet, despite these interesting developments in multilayered rhombohedral graphitic structures, experimental studies on bulk-size rhombohedral graphite are still scarce and much awaited.

Recently, much attention has been placed in understanding the different properties and signatures that different stacking orders produce in 2D materials [24,25]. In Raman scattering investigations, the characterization of the stacking can be extracted from the low-frequency shear modes, that provide information about the interlayer coupling [23]. Lui *et al.* [26] has compared the low-frequency Raman spectra of trilayer graphene with different stacking order, ABA and ABC, and have shown that they display considerably different shear mode signatures. Another approach within Raman experiments has been to probe the differences in electronic structures by means of multiple laser excitation characterization. Cong *et al.* [12] previously pointed out some of the differences in Raman features for trilayer graphene stacked in both Bernal and rhombohedral configurations, especially in the band widths and positions for the G, 2D and intermediate-frequency combination modes. [12]. Lui *et al*. [13] as well characterized three layers of ABA- and ABC-stacked graphene flakes by Raman spectroscopy, using four different laser lines with photon energies of 1.96 eV, 2.09 eV, 2.41 eV and 2.71 eV. More recently, Nguyen *et al*. [14] investigated the variation of the Raman features with the different laser excitation energies 1.96 eV, 2.33 eV, 2.41 eV, 2.54 eV and 2.81 eV for ABC-stacked graphene from 3 to 5 layers and different DR features can be used in order to distinguish both stacking configuration and the number of layers. Concerning bulk materials, the multitude of weak DR features have been investigated in Bernal graphite and interpreted in terms of combinations and overtones of finite momentum phonons from different branches. Contrarily to the case of few layer graphene, where the 2D and other DR features are fitted with a finite number of Lorentzians, the 2D band of graphite (2H and 3R) correspond to a convolution of contributions of phonons with momenta within the interior of the BZ, due to the dispersion that occurs along the $c^*$-direction. A complete theoretical calculation of the dispersive effects of the DR features in rhombohedral graphite is still lacking.

In the present work, we look towards elucidating the DR Raman scattering features of rhombohedral graphite. To this end, we perform Raman scattering on a mineral graphite sample employing different laser excitation energies in order to investigate the dispersion behavior of the DR bands and the resonance processes characterizing the graphitic structure produced by the ABC stacking order. The analysis of the 2D band of rhombohedral graphite revealed the existence of five double-resonance maxima, associated with resonance conditions involving different electronic branches. The other DR features involving acoustic phonons is also analyzed and their shape and dispersive behavior are compared to those of Bernal graphite.

## 2. Experimental

The investigated mineral graphite sample was obtained from Nacional do Grafite Ltda. and used as received. Raman spectroscopic characterization was performed at room temperature (~21 °C) employing two distinct systems: a WITec alpha 300R and a Horiba Jobin Yvon T64000. The former, which is equipped for acquiring spectra with the laser excitation energies 1.96 eV, 2.33 eV and 2.71eV, was used to measure Raman spectra at specific points on the sample as well as to perform confocal hyperspectral Raman imaging by means of its piezoelectric stage accessory. The latter was operated in the single-monochromator mode using an Innova 70C Ar-Kr as the CW laser source in order to obtain Raman spectra with the laser excitation energies at 1.92 eV, 2.18 eV, 2.41 eV and 2.54 eV. Spectra with the excitation energy at 2.81 eV were also collected in the triple-monochromator operation mode while employing a Kimmon He-Cd laser source.

All acquisitions were performed using a grating of 1800 grooves mm$^{-1}$, 100x objectives (NA 0.9) and incident laser power of 2 mW.

## 3. Results and Discussion

Fig. 1a presents an optical microscopic image of the region investigated by confocal hyperspectral Raman imaging within the graphitic sample. The image mainly focuses on two triangular regions: a large one and a small one; labeled respectively as R and B within the figure. Fig. 1b presents the Raman spectra in the 2D Raman band frequency region for these two investigated areas. Here, each spectrum represents the average profile obtained by Raman hyperspectral scanning and subsequently averaging the data within the respective regions. It is observed that the Raman spectrum of each region is markedly different: the 2D band widths and shapes differ, in Fig. 1b. Fig. 1c shows the Raman image that is formed when analyzing the spatial distribution of the 2D band width. From this image, one immediately notices that the distinct regions are indeed characterized by homogeneously distributed Raman spectra bearing distinct 2D band widths. These results are in accordance with the Raman fingerprints of ABC stacking for trilayer [12,13] and multilayer (~ 20 layers) graphene [16] and, therefore, indicate that, while the smaller triangular area is characterized as the Bernal stacking kind, the larger triangle consists of a graphitic structure with rhombohedral stacking order. Indeed, the large triangular area has lateral sides over 30 μm in length, yielding it an area of ~100 μm$^2$, and this relatively large size region represents a bulk rhombohedral sample with long-range ABC stacking order.

Fig. 1d additionally displays the hyperspectral Raman image generated by analyzing the spatial distribution of the $I_D/I_G$ ratio, *i.e.* the D band intensity relative to the G band one. This analysis allows one to map and pinpoint the structural defects within the investigated areas since the activation of the D band is dependent on the scattering from a defect center in order to conserve momentum [4,18]. From the Raman image, it may be seen that, although most of the regions are free from structural defects, two groups of edges are present: one being located at the interface of the Bernal and rhombohedral triangular regions and another one residing within the rhombohedral area itself. The spectra of the D band are provided in the Supplementary Information (SI) section (Fig. S1a), along with the details on the procedure of spectrum normalization.

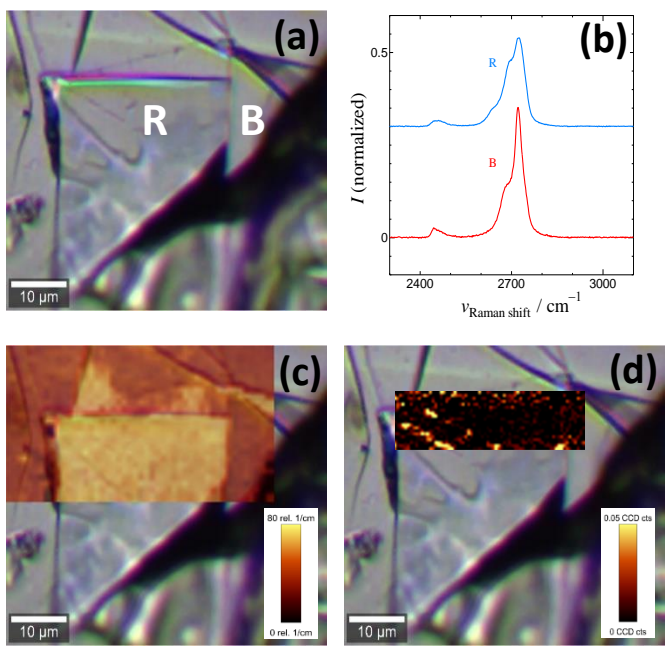

**Fig. 1 – (a) Optical microscopic image of the investigated area. Two triangular regions are respectively designated as R and B; (b) 2D Raman band spectra for R and B, obtained with laser excitation energy at 2.33 eV and extracted by averaging the hyperspectral data at the respective regions; (c) Overlaid Raman image representing the spatial distribution of the 2D band width measured with a laser excitation energy of 2.33 eV; and (d) Overlaid Raman image for the spatial distribution of the $I_D/I_G$ ratio collected with laser excitation energy at 1.96 eV.**

Fig. 2 exhibits the Raman spectra of Bernal and rhombohedral graphite performed with different laser lines, in the frequency region of the 2D band (the same figure without the overlaid fits is shown as Fig. S2 of the SI. The spectra were normalized respective to the prominent peak of the 2D band in Bernal-type graphite. In agreement with the studies on ABC-stacked few-layer [12–14] and multilayer [15,16] graphene, the 2D band of ABC graphite is wider and this character is seen to persist irrespective of the variation of the laser excitation energy.

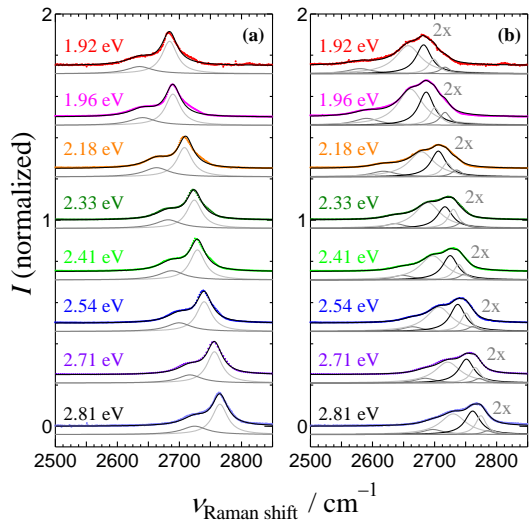

**Fig. 2 – The multiple-excitation Raman spectra of the 2D band region for graphite with ABA (a) and ABC (b) stacking orders. The ABC spectra was tentatively fitted with five Lorentzians: one corresponding to the main peak observed in the ABA stacking and four corresponding to processes I, II, III and IV, as identified from the spectral subtraction analysis (see Fig. 3). The Lorentzian components, shown below each spectrum, has been magnified in (b) by a factor of two for better visualization.**

The 2D band originates from an intervalley double-resonance (DR) Raman process involving phonons with different momenta within the Brillouin zone (BZ) [4], but its line shape in single-layer graphene is well described by a single Lorentzian curve. In the case of bilayer graphene stacked in the Bernal manner, the line shape basically involves four Lorentzian functions associated with the four allowed intervalley double-resonance processes, which occur due to the splitting of the valence and conduction bands into two parabolic sub-bands near the *K* point [4,27]. The line shape of the 2D band evolves with the number of graphene layers, reaching the signature of the bulk with around five graphene layers [28]. For bulk graphite, the 2D band line shape represents a sum of an infinitely large number of Lorentzian functions, since the DR process can occur at different layers along the c* axis of the three-dimensional BZ. This convolution is weighted by the phonon density of states (DOS) that satisfy the DR conditions along the c*-direction, which was shown to exhibit a singularity and a shoulder-like feature towards lower frequencies [27]. The 2D band of Bernal graphite can be fitted by two Lorentzian curves, as shown in Fig. 2a, but these peaks have no rigorous physical meaning. This fitting is only useful to get the position of the maxima in the phonon DOS that satisfy the DR process. The strong peak is due to the DR process along the line M-Γ-K in the BZ ($k_z = 0$) and the shoulder might come from processes along the line L-A-H in the BZ ($k_z = c^*/2$).

It is instructive to superimpose the 2D band spectrum of ABC graphite, obtained for each laser excitation energy, over that of ABA graphite (see Fig. S4 of the SI). By superimposing the spectra, one obtains a graph which roughly shows that the 2D band line shapes for ABC graphite present regions where they either overlap or slightly diverge from that of the ABA one. This, in turn, suggests that the different Raman features of the 2D band for the differently stacked graphite regions can be somewhat highlighted by performing subtraction of the ABA and ABC spectra collected under equal laser excitation energy conditions. Similar approach was previously undertaken by Torche *et al.* [16] – although the subtraction profiles were not actually shown – to deduce features of the long-range structure of rhombohedral graphite. In Fig. 3, we show the profiles obtained by directly subtracting the ABC spectrum from the ABA one (fitted carefully by means of multiple peak functions as to reproduce well the details of its line shape). The resultant profiles are essentially composed of positive (> 0) and negative (< 0) peak-like features; the positive ones being the remnant characteristics for the Raman spectra representative of the Bernal-type stacked graphite and the negative ones the intrinsic features present in the rhombohedral structure – presumably absent in the Bernal-type one –. Thus, the single peak that occurs on the positive side corresponds essentially to the high-frequency strong peak of the 2D band in ABA graphite. On the negative side of the profiles shown in Fig. 3, four processes are fundamentally identified: I, II, III and IV. These are seen to display dispersive behavior, since their positions vary with laser excitation energy; and therefore, we extracted the frequencies characterizing these processes from Fig. 3 by inspecting the first derivative curve of the respective subtraction profiles. This procedure is provided in Fig. S4, within the SI section.

In order to understand the origin of the four processes I-IV describe above, first notice that the primitive unit cell of rhombohedral graphite has only 2 carbon atoms [16], differently from the unit cell of Bernal graphite that has 4 carbon atoms. Therefore, the volume of the BZ of rhombohedral graphite is twice the volume of the Bernal graphite BZ. It is thus expected that other resonance channels can occur involving different electronic states within the BZ. This explains the fact that the 2D band of rhombohedral exhibit processes I to IV, related to four different maxima in the phonon DOS which satisfy the DR condition within the rhombohedral BZ.

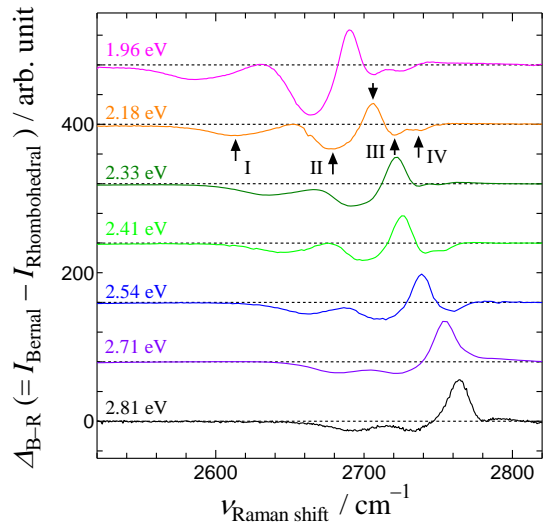

**Fig. 3 – Resultant profiles from the subtraction of the Raman spectra obtained for ABA and ABC stacking. The horizontal short-dashed line gives the zero baselines for each profile. The down-pointing arrow indicates the peak-like feature on the positive side, whereas the upwards-point arrows mark the processes occurring on the negative side. These processes are labeled by means of capital Roman numbers.**

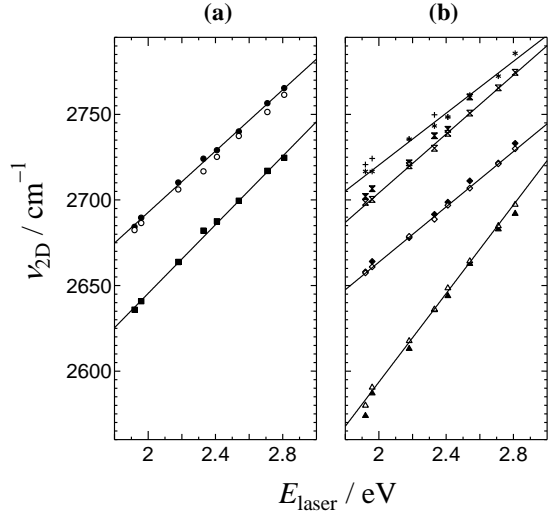

**Fig. 4 – Laser excitation-energy dependence of (a) the peak-like features identified in Fig. 3 and fitted in Fig. 2 for the Bernal-type structure: high-frequency peak (closed circles) and low-frequency shoulder (closed squares); and (b) the fitted peaks akin to the rhombohedral graphite: process I (triangles); process II (diamonds); process III (double triangles); and process IV (stars and crosses). In (b) the open and closed symbols, as well as the stars and crosses, correspond to the peak-tops as determined from Fig. 5 and 6, respectively. The open circles in (a) correspond to the additional Lorentzian peak used to fit the rhombohedral 2D spectra in Fig. 3. The linear fits in (b) were obtained considering only the fitted Lorentzian functions; and those in (a) only the closed symbols.**

In Fig. 2b, the 2D band line shapes of rhombohedral graphite are shown as fitted by means of five Lorentzian functions. These functions correspond to the four intrinsic processes depicted in Fig. 3 for rhombohedral graphite plus one which is analogous to the strong phonon DOS peak of Bernal graphite and designated here as process V. As in the case of Bernal graphite and given the 3D BZ in bulk materials, the line shape of the 2D in rhombohedral graphite is expected to reflect the phonon DOS that satisfy the DR condition. First-principles calculations of the phonon dispersion relations and the DR phonon DOS for both Bernal and rhombohedral graphite were carried out previously by Mounet and Marzari [29]; in their work slight differences were identified to exists only along the Γ-A direction and for some in-plane dispersions near Γ. Given these results, it seems plausible that the main maximum in the DR phonon DOS of Bernal graphite also exists for rhombohedral graphite. Therefore, this is the reasoning behind the inclusion of the additional fifth Lorentzian (process V) in Fig. 2b. The other maxima may be originated from DR processes that involving different electronic states and phonon branches in the rhombohedral BZ. To perform the fits in Fig. 2b, the centroids were fixed at the initial positions as determined from the analysis of Fig. 3 and the peak widths and heights of the four intrinsic processes were allowed to vary without restrictions. Subsequently, the peak widths were fixed at their mean values and then the peak centroids and heights adjusted to yield the optimal fitting curves. For the additional fifth Lorentzian, only the peak height and position were allowed to vary; the peak width being kept fixed at that as extracted from the graphite spectrum of the Bernal-type structure. Fig. 4 shows the dependence of the peak centroids on the laser excitation energy for all the processes characterized in Fig. 2 and 3.

Fig. 4a shows the dependence of the two peaks of the 2D band Bernal-type stacking structure on the excitation laser energy. The dispersive behavior of the strong peak is 89 cm$^{-1}$ eV$^{-1}$, being close to those of monolayer graphene and turbostratic graphite at 88 cm$^{-1}$ eV$^{-1}$ and 95 cm$^{-1}$ eV$^{-1}$, respectively [4,30]; and the slope of the shoulder is around 100 cm$^{-1}$ eV$^{-1}$.

Fig. 4b shows the dependence of the positions of processes I, II, III and IV observed in rhombohedral graphite as function of the excitation laser energy. Notice that the slopes of the two intermediate processes II and III, 83 and 89 cm$^{-1}$ eV$^{-1}$, respectively, are close to those of Bernal graphite. However, the slopes of processes I and IV of the rhombohedral structure are quite different, 132 and 61 cm$^{-1}$ eV$^{-1}$, respectively. This important difference may be ascribed to the occurrence of the new DR conditions involving other bands in the electronic structure of rhombohedral graphite. The slope values observed from our experiments are tabulated in Table 1, along with some reference ones.

Fig. 5 shows the variation of the integrated areas for the Lorentzian components used in Fig. 3 to fit the 2D band of the ABC graphite as a function of laser excitation energy. We can also present such areas as ratios to the full area of the 2D band, so that the percentage of contribution of each component could be evaluated. Such a representation is shown in Fig. S5 within the SI section, along with the full areas of the 2D band as a function of excitation energy for the different graphitic structures. Fig. 5 shows that the integrated intensities for most of the components used to fit the 2D band of ABC graphite practically do not depend on the excitation energy. Yet, a distinct behavior is clearly identified for process II, given its relatively large variation with laser excitation energy. This behavior thus indicates that – within our picture composed of 5 processes – the increase of the full 2D band area in rhombohedral graphite (see Fig. S5a) essentially results from the peak strength increase of process II upon decreasing excitation energy. This, in turn, strongly suggests that process II is a fundamental character akin to the rhombohedral structure. Indeed, it can be well observed from the previous multiple-excitation Raman studies on few-layer rhombohedral graphene that a common feature of the 2D band line shapes which distinctively characterize the ABC configuration from that of the ABA is that at low excitation energy they show a prominent peak-like feature appearing on the low-frequency side of the 2D band. This component presumably corresponds to process II, since the behavior depicted for it in Fig. 8c would yield a similar outcome. It is further worth noting that Torche *et al*. [16] has ascribed the presence of a shoulder-like feature at ~2576 $cm^{-1}$ in the Raman spectrum obtained with the laser excitation energy at 1.96 eV as a signature of long range rhombohedral order. In our data, this feature corresponds to process I and it is characterized by a relatively large slope in its dispersive behavior.

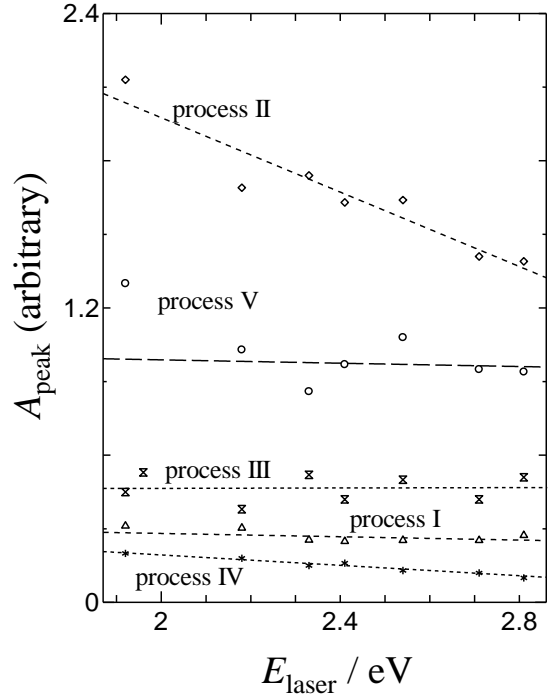

**Fig. 5** –The laser excitation-energy dependence for the Lorentzian integrated areas observed in rhombohedral graphite. The symbols respectively represent: process I (triangles); process II (diamonds); process III (double triangles); process IV (stars); and process V (circles).

Let us now consider the frequency region of the $G^*$ Raman band [4] (also referred in the literature and the D + D" band), shown in the multiple-excitation Raman spectra in Fig. 6a and 6b. Here too, differentiated band shapes are accordingly observed for the differently stacked configurations of graphite. The $G^*$ Raman band of Bernal graphite shown in Fig. 6a is asymmetric, exhibiting a maximum and a shoulder in the high frequency side. This band comes from an intervalley DR process involving one phonon of the iTO branch and one phonon of the LA branch, with opposite momenta near the K point. The asymmetric profile of the $G^*$ band comes from the fact that the frequencies of the iTO-LA and the LA-iTO processes are slightly different. The slope of the G* band dispersion for Bernal graphite is negative and around -34 cm$^{-1}$ eV$^{-1}$ in turbostratic graphite [30]. This result is due to the fact that the dispersion of the LA phonon branch is opposite to that of the TO branch, giving rise to the negative and relatively weak dispersive behavior of the G* band, when compared to the dispersion of the 2D band, for example. The G* band of rhombohedral appears to be broader and with less pronounced features. Such results are in accordance with the data reported for ABC trilayer graphene [14]. In Fig. 6, the G* band line shapes are fitted by means of two Lorenztian components: a narrow low-frequency peak and a broad high-frequency one. The dependence of the peak centroids on the excitation laser energy can be found in Fig. S6 and Table S1, within the SI section. The slopes of the dispersive behaviors of the high-frequency component of the G* band in Bernal and rhombohedral graphite are, respectively, -34 and -31 cm$^{-1}$ eV$^{-1}$, being, thus, about the same. Only the line shapes are different, which results from the decrease in peak intensity of the low-frequency component in rhombohedral graphite, as seen from Fig. 6.

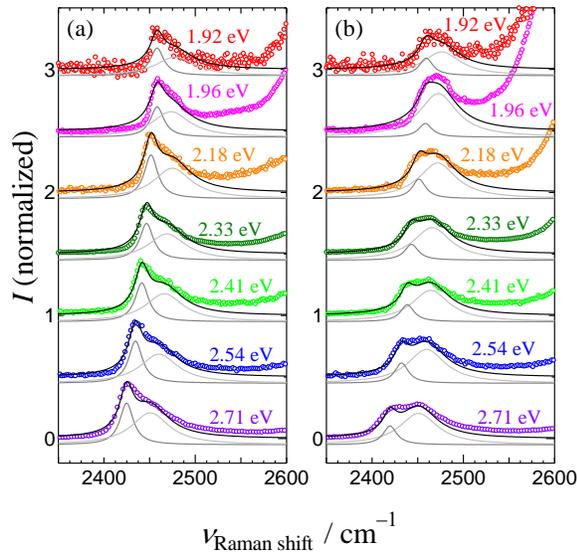

**Fig. 6 – Raman spectra of the 2D* band region for graphite with ABA (a) and ABC (b) stacking orders collected at different laser excitation energies.**

Fig. 7 presents the frequency range of the Raman spectra, between 1700 and 1850 cm$^{-1}$, where we can observe the so-called M bands. These bands, which are absent in the Raman spectra of monolayer graphene and poorly stacked graphene layers [31], show considerable different spectral features in few layer graphene (3 to 5 layers) according to the type of stacking configuration [14]. In Fig. 7, such differences are confirmed to hold for the bulk graphite as well. The M bands come from an intravalley DR process, involving one optical and one acoustic phonon near the Γ point. Sato *et al.* [32] studied such modes in bilayer graphene, while Cong *et al*. investigated them in the trilayer structures ABA and ABC [12]. They are ascribed to the first overtone of the oTO phonon mode, *i.e.* 2oTO, and combination modes of longitudinal (LO) and out-of-plane (ZO) phonon modes, *i.e.* LOZO. Calculations predicted that the LOZO modes show positive dispersion with laser excitation energy, while the 2oTO (q = 2k) modes display negative dispersive behaviors [12,32]. In Fig. 7, the spectra have been fitted with several Lorentzian functions to the extent that the measurement resolution allows; the dispersive behaviors of the singled-out peaks are portrayed in Fig. 8. On one hand, inspection of the dispersive behaviors of the Lorentzian components for ABA graphite in Fig. 7a allows straightforward identification of five M modes: two exhibiting negative dispersions and three showing positive dispersions. These can be ascribed to 2oTO and LOZO modes, respectively, based on the characters of their dispersion behaviors. On the other hand, in the case of ABC graphite, the bands appear within a smaller frequency window and their experimental distinction – and consequently ascription – becomes harder. Nevertheless, in Fig. 7b Lorentzian centroids have been tentatively grouped up by identifying dispersive behaviors that would yield similar slopes as those extracted for the modes in the ABA structure. For verification, the slope values are given within Table S2, found within the SI section. This analysis suggests that the 2oTO modes blueshift (~ 2–8 $cm^{-1}$) and the LOZO modes redshift (~ 10–20 $cm^{-1}$) in the ABC structure, as compared to the ABA one. It is worth noting that other combination modes, such as iTA+iTO, iTA+LO, iTO+LA and LO+LA, occur within the frequency window above ~ 1800 $cm^{-1}$ in bilayer graphene [32], ABA and ABC structures of trilayer graphene [12] and Bernal graphite [33]; and thus their existence are unlikely to affect the analysis presented in Fig. 7b.

In Fig. 7, some of the Lorentzian components of the corresponding modes within the differently stacked structures are highlighted by means of matching colors for their easy visualization. It is seen that the peak intensity for the highest-in-frequency LOZO mode, *i.e.* LOZO(1), diminishes considerably when going from the ABA structure to the ABC one. Such behavior presumably reflects an unfavorable change in the DR process of the ABA structure. The number of features is greater for the ABC structure, a result that can be also ascribed to the doubling of the BZ volume in the ABC structure compared to the ABA structure, which ultimately lead to the creation of extra channels for the DR processes.

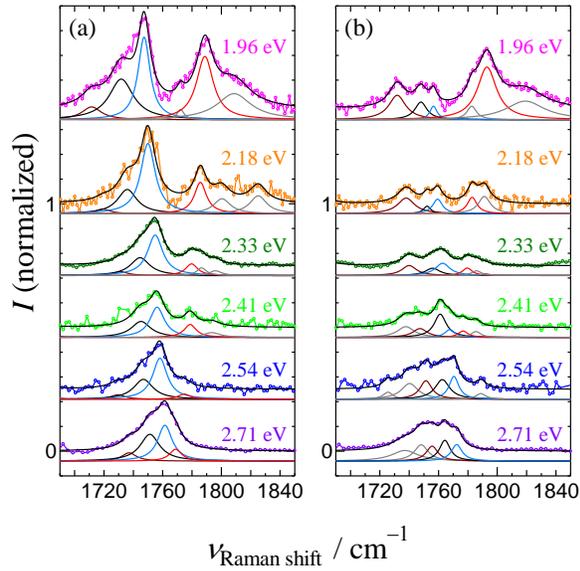

**Fig. 7 – The Raman spectra of the intermediate-frequency region (M bands' region) for bulk graphitic structures with (a) ABA and (b) ABC stackings, collected at different laser excitation energies. The spectra were fitted with multiple Lorentzian functions; and the components are exhibited underneath each spectrum, shifted vertically. The corresponding bands in both structures, tentatively identified from the analysis of the dispersive behaviors of the components (see Fig. 8), are matched by applying different colors to the Lorentzian components.**

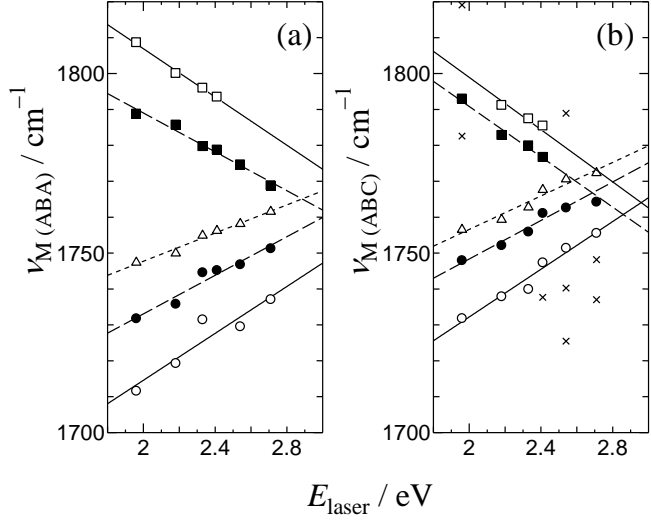

Fig. 8 – The dispersive behaviors of the M bands for graphite with ABA (a) and ABC (b) stacking structures. The M modes occurring with q = 2k are ascribed as following: 2oTO(3) (open squares); 2oTO(1) or/and (2) (closed squares); LOZO(1) (open triangles); LOZO(2) (closed circles); and LOZO(3) (open circles). The data points plotted with the x marks represent modes that have not been ascribed.

## 5. Concluding remarks

We have performed in this work a multiple-excitation Raman study of rhombohedral graphite using eight different excitation laser energies in the visible range, that allowed us to obtain the slope of the dispersion of different double-resonance (DR) Raman bands. All results were compared to those of Bernal graphite, and analyzed in terms of the DR Raman processes within the rhombohedral Brillouin zone (BZ). By subtracting the Raman spectra of the two different stacked types of graphite, it was possible to evidence the features of the 2D band of rhombohedral graphite, and the results show that it is broad since it is composed by five distinct processes which exhibit quite different dispersive behavior. The double resonance spectra involving combination of different optical and acoustic phonon branches, such as the G* band around 2450 cm$^{-1}$ and the M band in the range 1700-1850 cm$^{-1}$, were also measured and analyzed by decomposing the bands, and their dispersive slopes were obtained from the multiple-excitation Raman spectra. Our results are interpreted in terms of the existence of new channels for the DR process in rhombohedral graphite when compared to Bernal graphite since the volume of its BZ is twice that the later, allowing other resonance conditions involving different states within the BZ. Our results will be useful for future calculations of the DR spectra of rhombohedral graphite that, in turn, will allow the precise assignment of the different DR features and their dispersive behavior in rhombohedral graphite.

**Table 1 – Slopes of the various dispersive behaviors observed for second order Raman bands in rhombohedral (ABC) and Bernal-type (ABA) graphitic structures.**

| Rhombohedral | $\alpha_{disp}$ (cm$^{-1}$ eV$^{-1}$) | Bernal-type | $\alpha_{disp}$ (cm$^{-1}$ eV$^{-1}$) |
|---|---|---|---|
| **D Raman band** | | **D band** | |
| | 52 | | ~50[†] |
| **2D Raman band** | | **2D Raman band** | |
| *Process I* | 132 | *Main Process* | 90 |
| *Process II* | 86 | *Secondary Process* | 100 |
| *Process III* | 84 | | |
| *Process IV* | 78 | | |

[†] ref. [4]

**Acknowledgements**

This work was partially supported by the Brazilian Institute of Science and Technology (INCT) in Carbon Nanomaterials and the Brazilian agencies Fapemig, CAPES and CNPq. Professor Ana Champi is grateful to 2MI and VRS Eireli Research & Development for their financial support. Finally, many thanks to CTNano-UFMG, for providing use of equipaments, and to Nacional do Grafite – MG, for providing samples of high quality.

**Supplementary Information: Multiple-excitation study of the double-resonance Raman bands in rhombohedral graphite**

Sergio L. L. M. Ramos[1,†], Marcos A. Pimenta[1], Ana Champi[3,*]

[1] *Departamento de Física, Universidade Federal de Minas Gerais (UFMG), Belo Horizonte, Minas Gerais 30123-970, Brazil*

[2] *Centro de Ciências Naturais e Humanas, Universidade Federal do ABC (UFABC), Santo André, São Paulo 09210-170, Brazil*

[†] present address: *School of Chemistry, University of Manchester, Manchester M13 9PL, United Kingdom*

* Corresponding author. E-mail: ana.champi@ufabc.edu.br

**Figure and table contents:**

- Fig. S1: Raman spectra of the D and the D'-overtone bands for rhombohedral graphite.

- Fig. S2: Raman spectra of the 2D band region for graphite with ABA and ABC stacking orders collected at various laser excitation energies.

- Fig. S3: Superimposed Raman spectra of 2D bands for ABA and ABC graphite at various laser excitation energies.

- Fig. S4: First-order derivatives of the profiles obtained from the spectral subtraction analysis.

- Fig. S5: Integrated areas of the 2D bands for ABA and ABC graphite and of the Lorentzian components composing the 2D band of rhombohedral graphite.

- Fig. S6: Laser excitation-energy dependence of the Lorentzian components composing the $G^*$ band in ABA and ABC graphite.

- Table S1: Slopes of the dispersive behaviors observed for the G$^*$ Raman bands in graphene, ABA graphite and ABC graphite.

- Table S2: Slopes of the various dispersive behaviors observed for Raman bands identified within the intermediate-frequency region (M bands) in ABA and ABC graphite.

**Spectral normalization and subtraction analysis**

The Raman spectra of Bernal graphite was normalized to the height of the 2D band, following the subtraction of the background contributions by employing second order polynomials. To accomplish equivalent normalization for the spectra of rhombohedral graphite exhibited in Fig. 2b, analogous background subtraction was first performed and, subsequently, the G bands heights were adjusted as to overlay and respectively match the line shapes of those collected for Bernal-type graphite. The G peak centroids were also carefully aligned, neglecting any potential redshift (~0.5 cm$^{-1}$) that could be expected in rhombohedral graphite [Cong, C.; Yu, T.; Sato, K.; Shang, J.; Saito, R.; Dresselhaus, G. F.; Dresselhaus, M. S. Raman Characterization of ABA- and ABC-Stacked Trilayer Graphene. *ACS Nano* **2011**, *5* (11), 8760–8768. https://doi.org/10.1021/nn203472f].

To reveal the spectral features of the Raman 2D band for rhombohedral graphite, the spectra of Bernal graphite was subtracted from that of the rhombohedral one. To this end, the 2D band of the normalized Bernal graphite spectra was arbitrarily fitted by four Lorentzian peak functions, as to reproduce well its line shapes. Then, the intensity values at the respective frequencies were computed and directly subtracted from each data point of the normalized rhombohedral spectra. This resulted in spectral subtraction profiles with a clear systematic behavior. Nevertheless, it was then realized that for a few spectra, such as that acquired at 1.96 eV, for example, the visualization of the systematic behavior could be better aligned to the others by multiplying additional weight factors to the Bernal graphite spectra during the subtracting of spectra. Under this consideration, a few of the profiles displayed in Fig. 3 were produced by employing such weight factors.

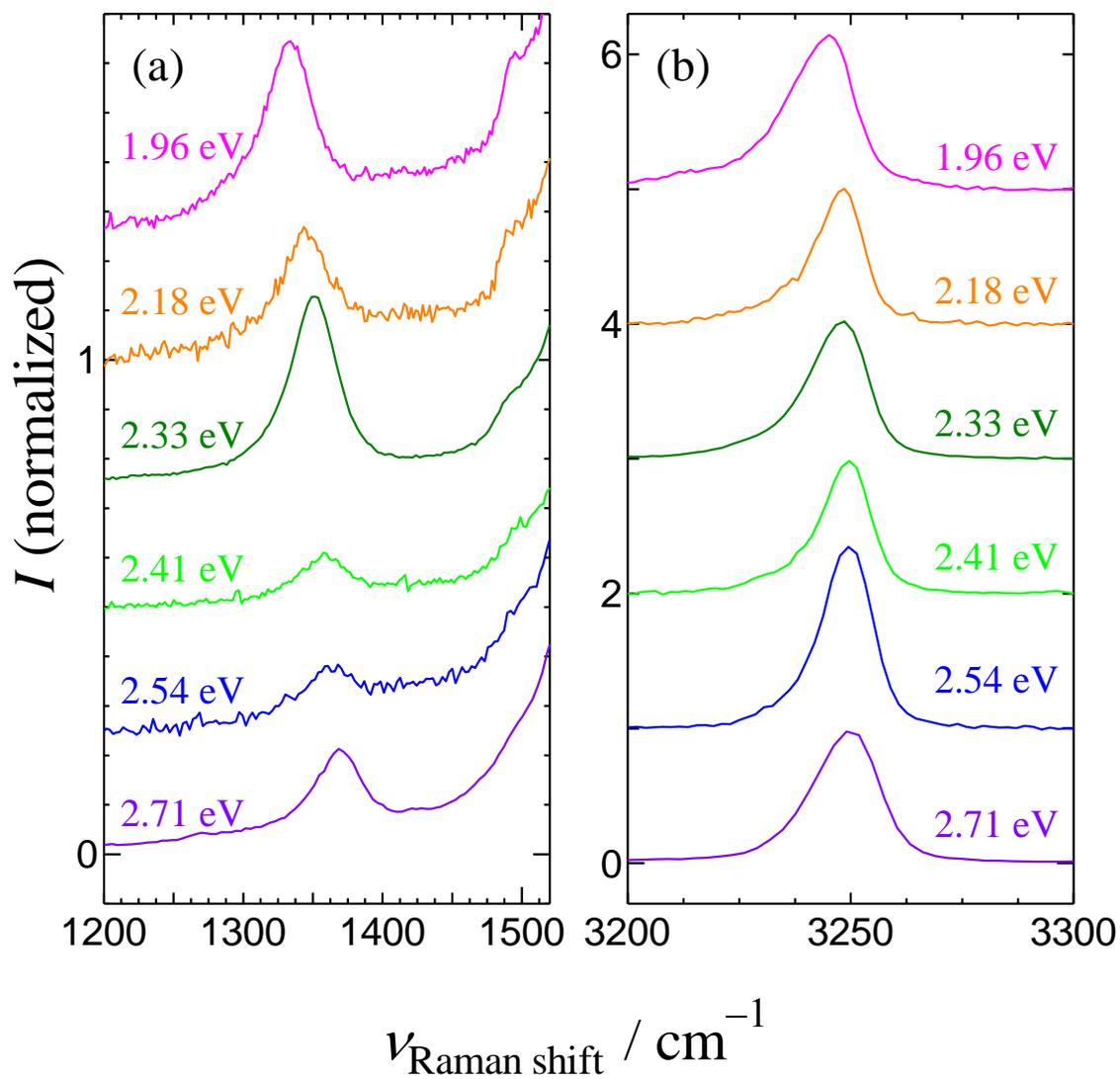

**Fig. S1** – Raman spectra of the D (a) and the D'-overtone (b) bands for rhombohedral graphite collected at various laser excitation energies.

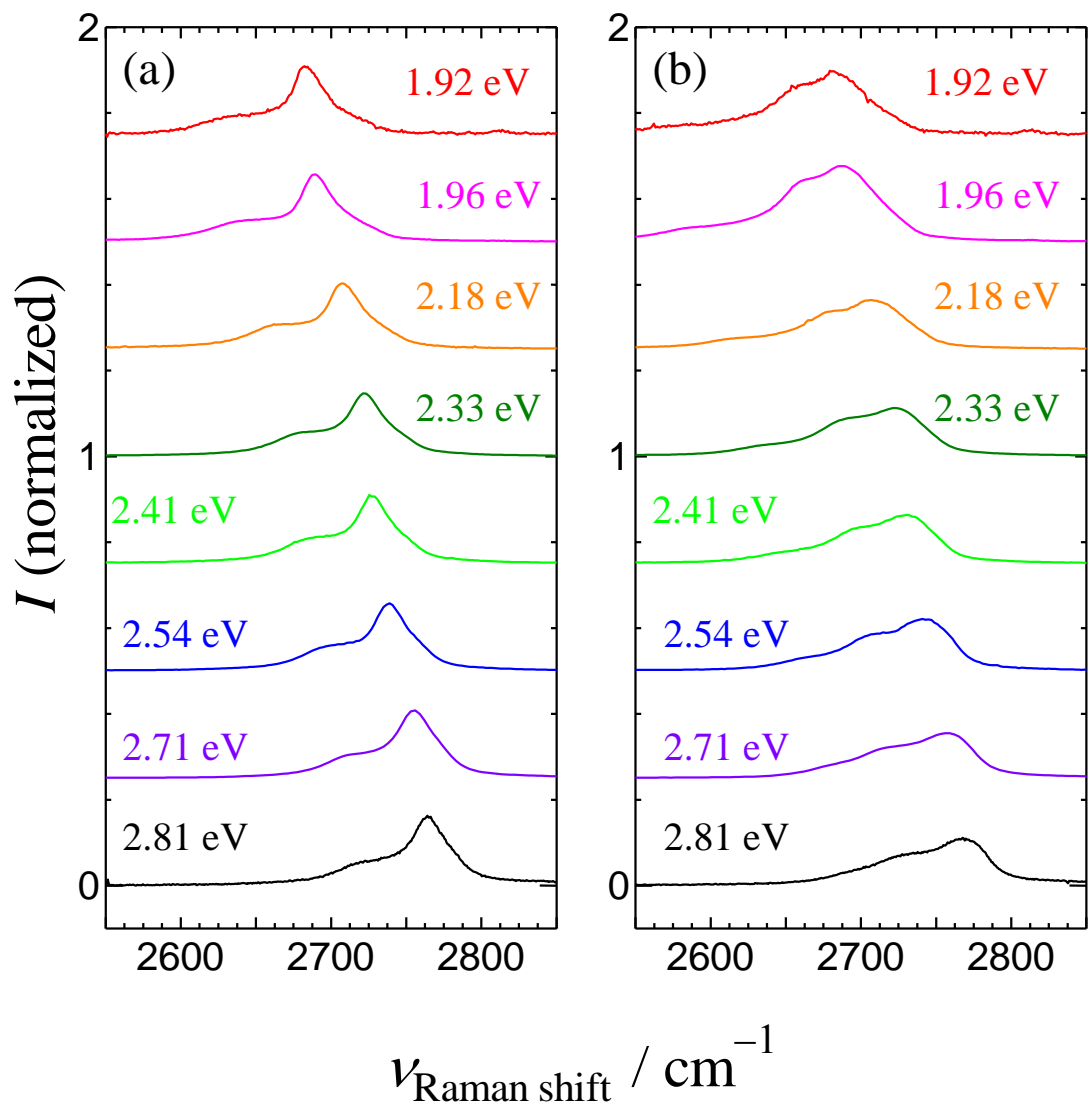

**Fig. S2** – Raman spectra of the 2D band region for graphite with ABA (a) and ABC (b) stacking orders collected at various laser excitation energies.

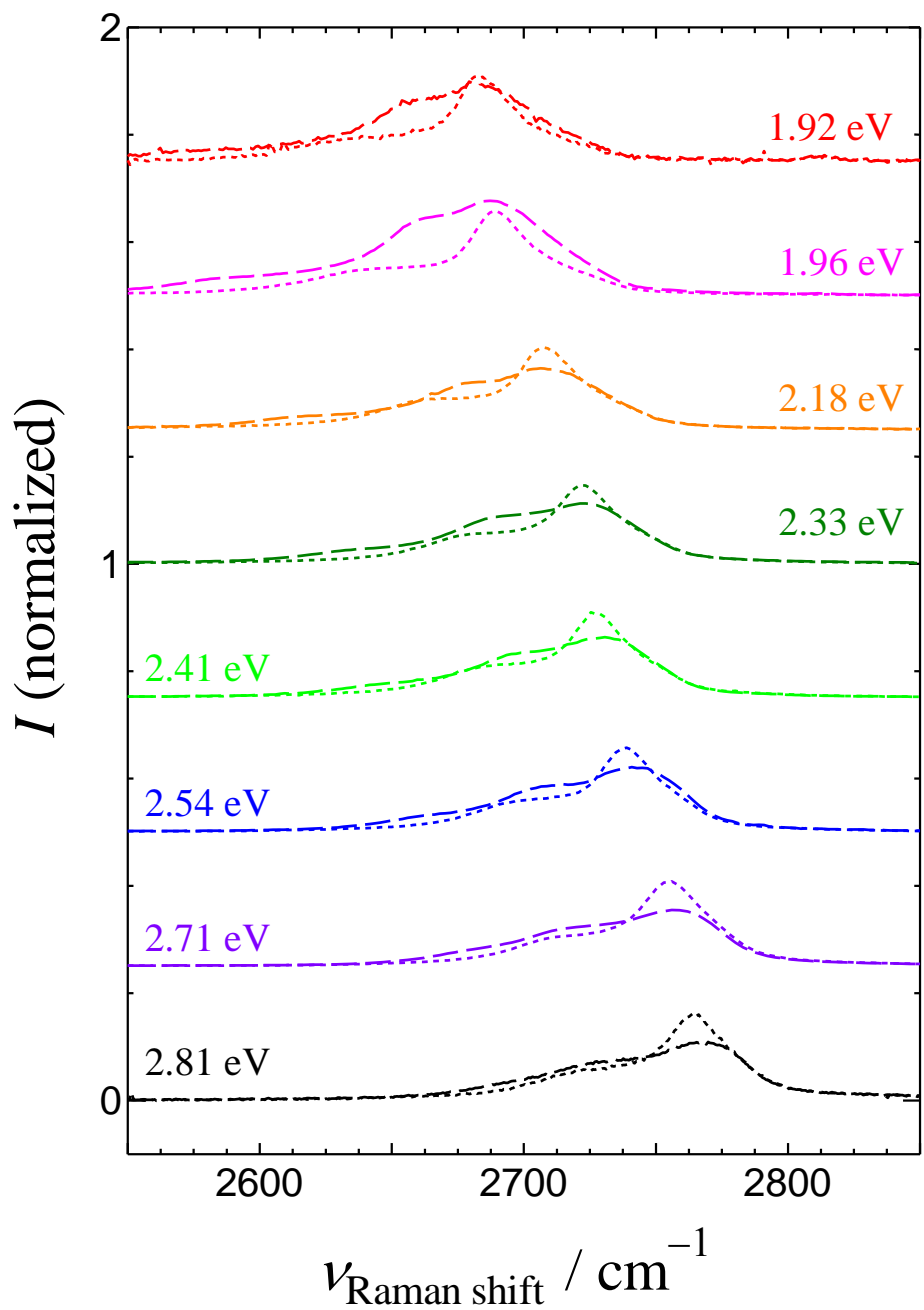

**Fig. S3** – Superimposed 2D Raman spectra of ABA (dotted lines) and ABC (dashed lines) at the various laser excitation energies.

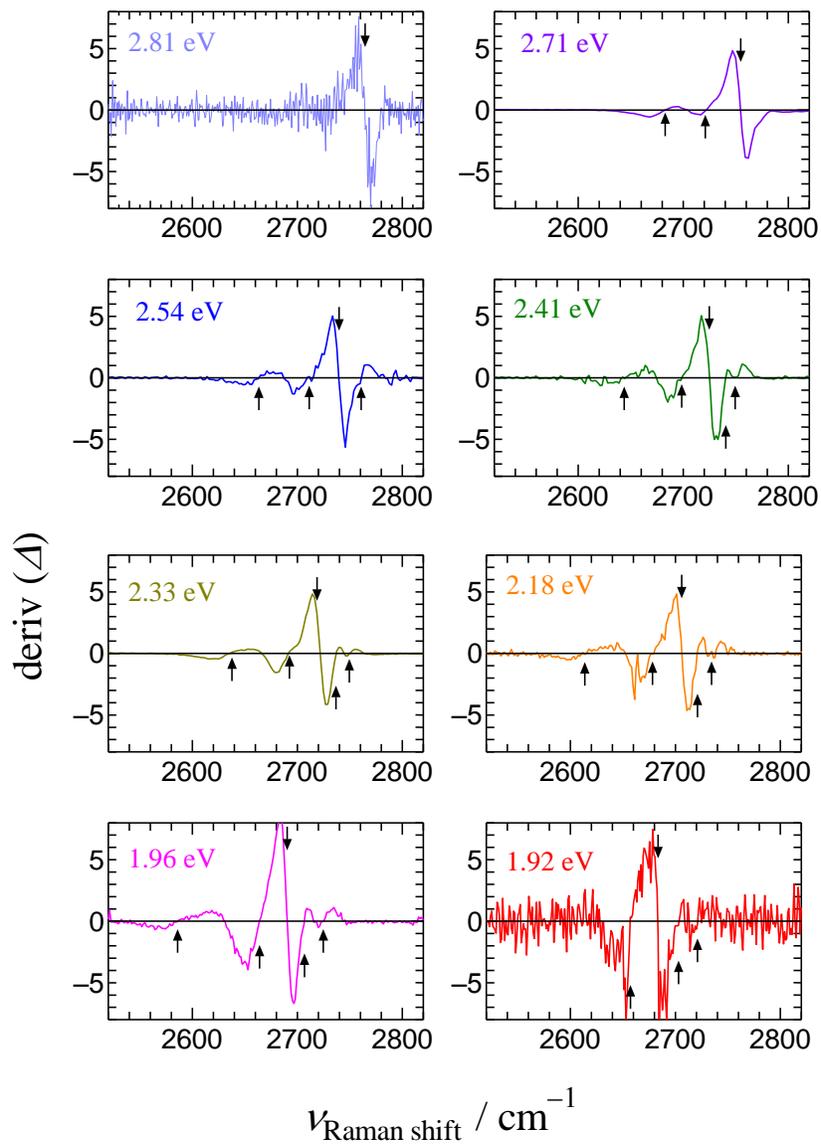

**Fig. S4** – First-order derivatives of the profiles obtained from the spectral subtraction analysis displayed in Fig. 3. The arrows indicate the points at which the derivative is null, indicating a peak maximum.

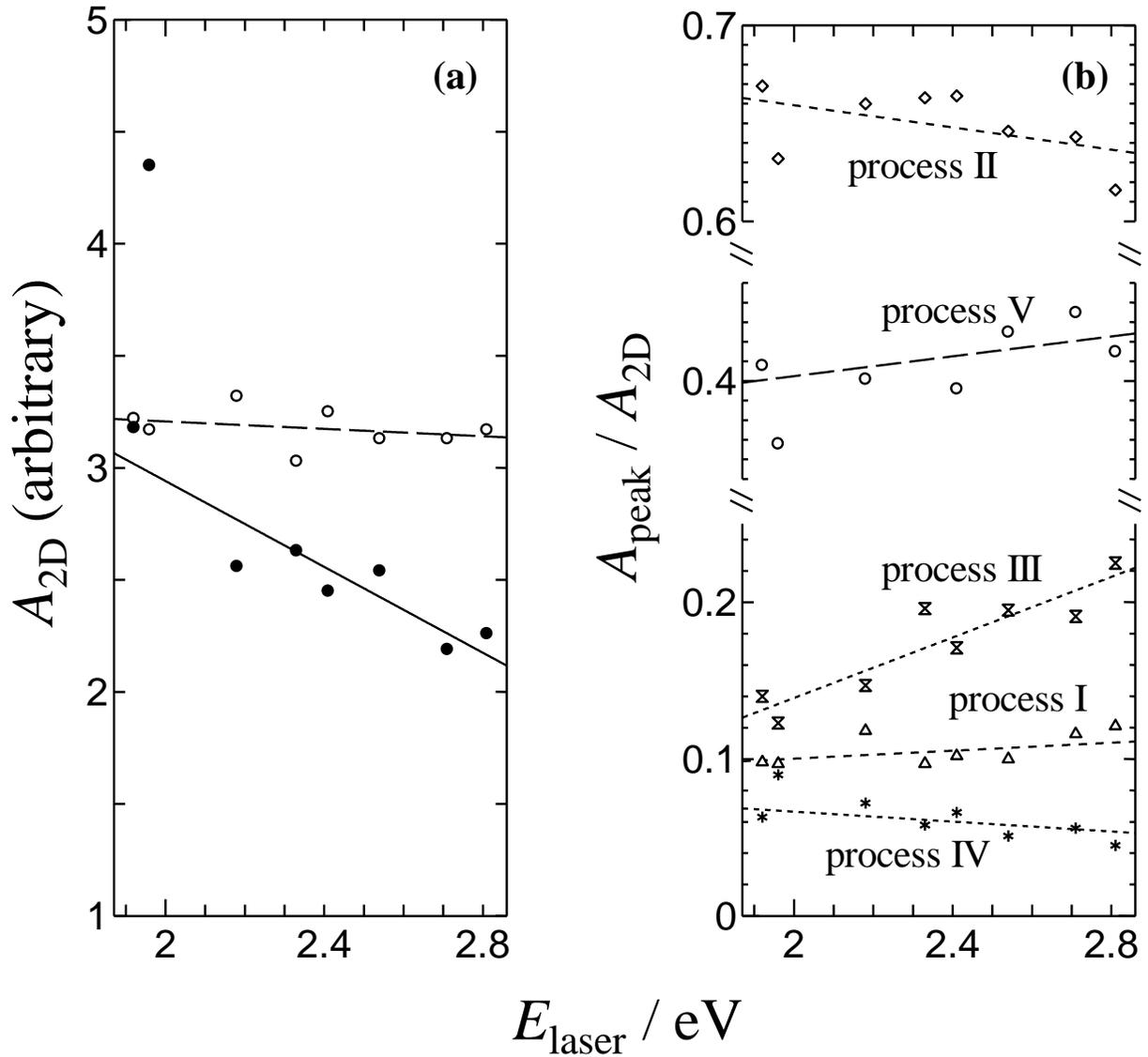

**Fig. S5** – The laser excitation-energy dependence of (a) the total area of the 2D band for Bernal (open circles) and rhombohedral (closed circles) graphite; and (b) the ratios of the Lorentzian integrated areas, as normalized by the total area of the 2D band, observed in rhombohedral graphite. The symbols respectively represent: process I (triangles); process II (diamonds); process III (double triangles); process IV (stars); and process V (circles).

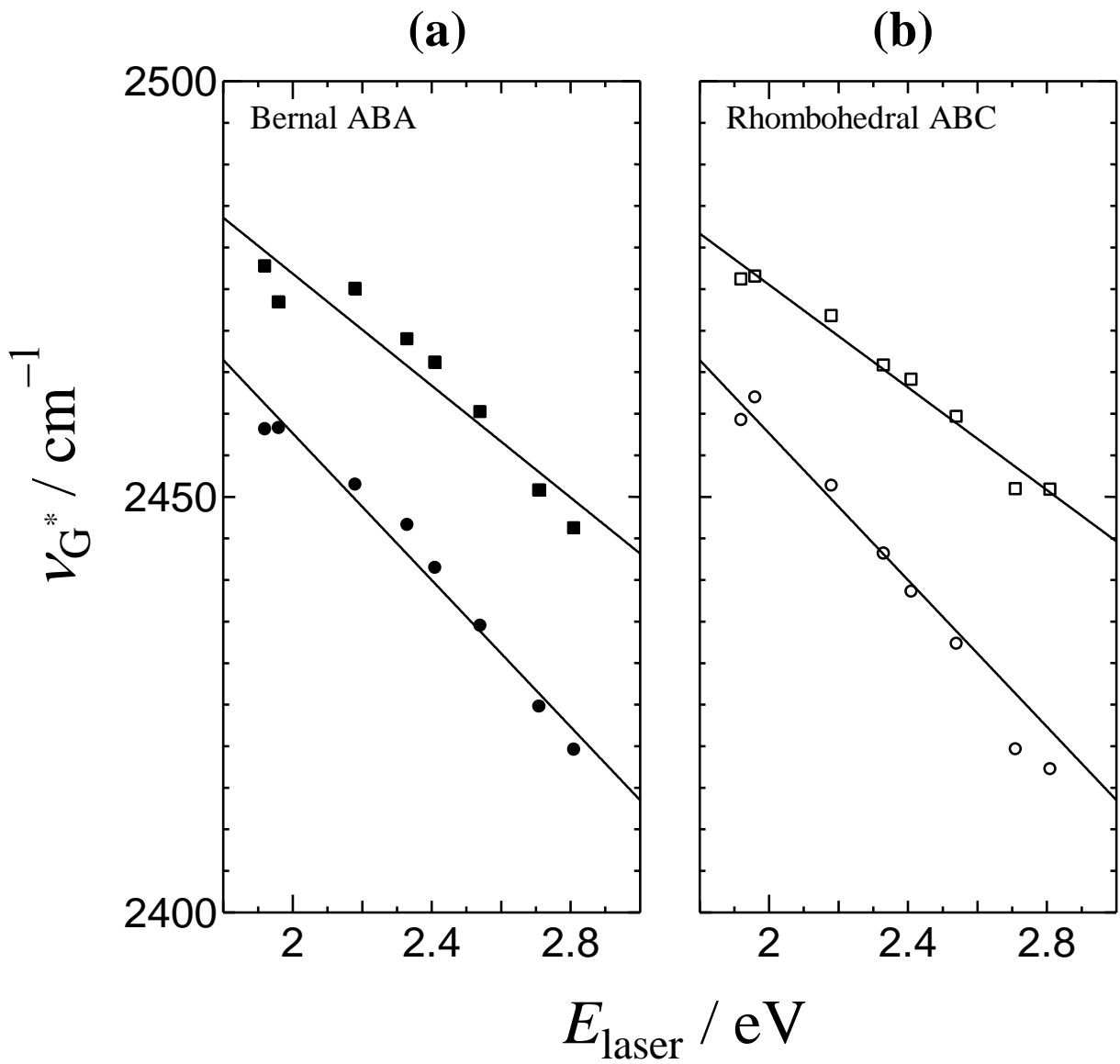

**Fig. S6** – Laser excitation-energy dependence of the two peak components for the G* band in Bernal-type graphite (left) and rhombohedral graphite (right). The low-frequency component is represented by circles, while the high-frequency one by square.

**Table S1** – Slopes of the dispersive behaviors observed for the G* Raman bands in graphene, Bernal (ABA) graphite and rhombohedral (ABC) graphite.

|  | $\alpha_{disp}$ (cm$^{-1}$ eV$^{-1}$) |
|---|---|
| **Monolayer graphene** | -18[†] |
| **Bernal turbostratic graphite** | -31[†] |
| **Bernal graphite** |  |
| *High-frequency* | -34 |
| *Low-frequecy* | -44 |
| **Rhombohedral graphite** |  |
| *High-frequency* | -31 |
| *Low-frequecy* | -44 |

[†] Ref. [Mafra, D. L.; Samsonidze, G.; Malard, L. M.; Elias, D. C.; Brant, J. C.; Plentz, F.; Alves, E. S.; Pimenta, M. A. Determination of La and to Phonon Dispersion Relations of Graphene near the Dirac Point by Double Resonance Raman Scattering. *Phys. Rev. B - Condens. Matter Mater. Phys.* **2007**, *76* (23), 3–6. https://doi.org/10.1103/PhysRevB.76.233407]

**Table S2** – Slopes of the various dispersive behaviors observed for Raman bands identified within the intermediate-frequency region (M bands) in Bernal-type (ABA) and rhombohedral (ABC) graphite.

| **Bernal-type** | | **Rhombohedral** | |
|---|---|---|---|
| | $\alpha_{disp}$ (cm$^{-1}$ eV$^{-1}$) | | $\alpha_{disp}$ (cm$^{-1}$ eV$^{-1}$) |
| **Raman M bands** | | **Raman M bands** | |
| 2oTO (3) | -34 | 2oTO (3) | -36 |
| 2oTO (1) or/and (2) | -31 | 2oTO (1) or/and (2) | -35 |
| LOZO (1) | 20 | LOZO (1) | 23 |
| LOZO (2) | 27 | LOZO (2) | 27 |
| LOZO (3) | 33 | LOZO (3) | 33 |